\documentclass[preprintnumbers,amsmath,amssymb,floatfix,11pt,prd,onecolumn,
superscriptaddress,nofootinbib]{revtex4}
\pdfoutput=1
\usepackage{latexsym}
\usepackage{epsfig}
\usepackage{epstopdf}
\usepackage{amssymb}
\begin{document}

\title{\bf Products of Thermodynamic Parameters for the Generalized Charged Rotating Black Hole and the Reissner-Nordstr\"{o}m Black Hole with Global Monopole}
\author{Bushra Majeed}
\email{bushra.majeed@sns.nust.edu.pk}
\affiliation{School of Natural
Sciences (SNS), National University of Sciences and Technology
(NUST), H-12, Islamabad, Pakistan}

\author{Mubasher Jamil}
\email{mjamil@sns.nust.edu.pk}\affiliation{School of Natural Sciences (SNS),
National University of Sciences and Technology (NUST), H-12,
Islamabad, Pakistan}

\begin{abstract}
{\bf Abstract:} We investigate the thermodynamics of Kerr-Newman-Kasuya black hole and the Reissner-Nordstr\"{o}m black hole with a global
monopole  on inner and outer horizons. 
Products of surface gravities, surface temperatures, Komar energies, electromagnetic potentials, angular velocities,
areas, entropies, horizon radii and the irreducible masses at the
Cauchy and the event horizons are calculated. It is observed that the product of
surface gravities, surface temperature product and product of
Komar energies, electromagnetic potentials and angular velocities at horizons are not universal quantities for these  black holes. Products of areas and
entropies at horizons are independent of masses of black holes.
 Heat capacity is calculated for the generalized charged rotating black hole and phase transition is observed, under certain conditions on $r$.
\end{abstract}
\maketitle
\newpage
\section{Introduction}The most interesting objects in theoretical physics are arguably black holes. To understand
their dynamics we need to put together two widely accepted theories of Nature: general
relativity (Einstein's classical theory of gravity) and quantum mechanics. Black hole thermodynamics
is the crossroad between the classical and the quantum pictures. Discovery of Hawking radiations lead to the identification
of black holes as thermodynamic objects with physical temperature and the  entropy  \cite{ch9670,ch5271,pe7771,be3373}.
This discovery  paved
the way for progress in the understanding of spacetime, quantum mechanically \cite{ha3074,ha9975}.
 Variation in the mass, ${\cal M}$, of a rotating black hole having angular momentum, $J$, and electric charge $Q$, obeys the formalism
\begin{equation}d{\cal M}= T d{\cal S}+\Omega d {J}+ \Phi^Q d{Q}\label{k0},\end{equation}known as first law of thermodynamics. Here  $\Omega$ is the angular velocity
of the horizon, $\Phi^Q$ is the electric potential on the horizon, due to the  electric charge $Q$,  and ${\cal S}$ is the entropy of the black hole.
Important results of the black hole thermodynamics are the association of temperature ($T$) and the entropy $({\cal S})$ with
the  surface gravity $(\kappa)$ and  the area $({\cal A})$ of the black hole event horizon respectively. 
The phenomena of phase transition in the  black hole thermodynamics
was first observed long ago \cite{pa9590,ne1865}. The Schwarzschild black hole has negative specific heat and evaporates via Hawking
radiation. The AdS Schwarzschild black holes have a different behavior towards temperature and heat capacity. There are two types of black holes in AdS spacetime:  the smaller one, like the usual Schwarzschild black hole, with
negative specific heat (unstable), and big black holes having positive specific heat (locally stable).
Axisymmetric, stationary, and
electrically charged black holes in Einstein-Maxwell theory with arbitrary surrounding matter,
always have regular inner horizon (Cauchy horizon, ${\cal H}^-$) and
an outer horizon (event horizon, ${\cal H}^+$), if the angular momentum and charge of the black hole do not vanish at the same time \cite{an0108}.
In recent years products of thermodynamics parameters, specially  the area and the entropy, at both horizons of black holes has gained attention in general relativity and string theory \cite{cv0111}.
It is observed that the area product of the outer
and inner horizons is independent of the  black hole mass ${\cal M}$. For a regular axisymmetric and stationary spacetime in Einstein Maxwell gravity these products are \cite{wi0598}:
\begin{equation} \mathcal{A}_+\mathcal{A}_-=(8\pi)^2( J^2+\frac{Q^4}{4}),\end{equation}
and \begin{equation}{\cal S_+}{\cal S_-}= (2\pi)^2( J^2+\frac{Q^4}{4}).\end{equation}
Note that both the above given products are mass independent so these are universal quantities. This universal sense of  the area and entropy products holds
for all known five dimensional asymptotically flat black rings, and for black strings \cite{ca0812}.
Microscopic degrees of freedom
of the black hole are described in terms of those
of a conformal field theory (CFT). The area product of the inner and outer
horizons of a black hole in three dimensions is
\begin{equation}\frac{ \mathcal{A}_+\mathcal{A}_-}{(8 \pi G_3)^2}=N_R-N_L, \label{2d}\end{equation}
where $N_R$ and $N_L$ are the number of right and
left moving excitations of the two-dimensional  CFT \cite{fl0597, ca0812}, i.e.
\begin{equation}\frac{\mathcal{A}_+\mathcal{A}_-}{(8 \pi G_d)^2} \in \mathbb{Z}.\label{1d}\end{equation} In other words, products
of areas of Killing horizons  is independent of the mass of any asymptotically
flat black hole in d-dimensional spacetime, therefore depends on the quantized charges \cite{ca0812}.
Hence mass independence of  the area products, is necessary condition for the  holographic CFT description.
In \cite{cv0111, pr8714, prarx14} thermal products for rotating black holes are studied. In \cite{vi1413} area products for stationary black hole horizons are calculated.
It has been shown that area products are
independent of the ADM (Arnowitt-Deser-Misner) mass parameter and depend on the quantized charge and quantized angular momentum parameter
for all known five dimensional asymptotically flat black rings and  black strings. It may sometimes also fail, e.g. in \cite{li5410} authors show that entropy products are not mass independent in general Myers-Perry black holes when  spacetime dimension $d\geq 6$, and Kerr-AdS black holes with $d\geq 4$.
The Kerr/CFT correspondence from the thermodynamics of both outer and inner horizons  was investigated recently \cite{ch1712}. Authors prove that the first
law of thermodynamics of the outer horizon guaranty that of the inner horizon, under some assumption, and  mass independence of the entropy product ${\cal S}_+{\cal S}_-$ is
equivalent to the condition $ T_+ {\cal S_+} ={T}_-{\cal S_-} $. Furthermore, using the thermodynamics method, information of
the dual CFT could be obtained easily, because thermodynamics of the outer and inner horizons give the thermodynamics in the left and right moving sectors of the dual CFT \cite{cv0909}. Therefore, central charges and temperatures in all possible pictures can be obtained in a simple way.

Newman et. al. obtained the solution of Einstein-Maxwell equations in
the Kerr spacetime, as a rotating ring of
mass and electric charge \cite{ne1865}. Applying the Ernst's
formulation (for axisymmetric stationary fields), Tomimatsu and Sato
discovered the series of solutions for the gravitational
field of a rotating mass
 \cite{to4472,er7568}. Yamazaki obtained the charged
Kerr-Tomimatsu-Sato family of solutions with some  distortion parameter $\delta$ (integer) in the gravitational
fields of rotating masses \cite{ya0277}.
Static spherically symmetric
Julia-Zee dyon solution in curved spacetime were obtained by Kasuya et. al. \cite{ka181}.
Later, an exact stationary
rotating dyon solution in Tomimatsu-Sato-
Yamazaki spacetime, was proposed \cite{ka5181}. Furthermore, both the
Schwinger and the Julia-Zee, dyon exact solutions
in the the Kerr-Newman spacetime, i.e. for $\delta=1$, were studied. This solution is known as the Kerr-Newman-Kasuya (KNK) black hole in literature \cite{ka482}.
This solution is featured
by four physical parameters (mass ${\cal M}$, angular momentum $J$, electric charge $Q$, and magnetic
charge $P$).
 In this
work we consider the KNK black hole  and discuss its thermodynamic at horizons and generalize some
already existing results for the  Cauchy horizon ${\cal H}^-$.

Further we will extend these calculations to investigate the thermodynamics properties of  a charged, static spherically symmetric black hole with a global
monopole. This black hole is topologically quite different from that of the Reissner-Nordst\"{o}m black hole
because the
background spacetime is not considered asymptotically flat here, and it
contains a topological defect due to the presence of a global monopole \cite{ba4389}

Plan of the work is as follows:
In section (II), we discuss the KNK black hole metric and  products of quantities, surface
gravity, temperature and Komar energy, angular velocities and electromagnetic potentials,  the area and the entropy, are defined at  the inner and outer horizons, $\mathcal{H}^{\pm}$ . In section (III) the first law of thermodynamics is studied, we also get this law from  the Smarr formula
for the KNK black hole, the
irreducible mass of the black hole is also studied there. In section (IV), heat capacity of the black hole is calculated, and the phase transition is discussed. In section (V) metric of the Reissner-Nordstr\"{o}m black hole with the global monopole, (RN-GM), is discussed and   products of thermodynamics parameters are calculated on the inner and outer horizons. In section (VI) the first law of thermodynamics is discussed on horizons of the RN-GM black hole. In (VII) the first law of thermodynamics is obtained for the RN-GM black hole by applying the Smarr formula. In the last section concluding remarks are given. We use the units with $G = \hbar = c = 1$.

\section{The Kerr- Newman-Kasuya Black Hole}
The spherically asymmetric and rotating solutions for Einstein-Maxwell field equations, in the presence of electric and magnetic field, was proposed by Kasuya
 \cite{ka482}. Metric in the Boyer- Lindquist coordinates is
\begin{eqnarray}\label{k1}
ds^2&=& -\Big(1-\frac{2{\cal M}r-Q^2-P^2}{\Sigma}\Big)dt^2 - \frac{2a\sin^2{\theta}}{\Sigma}(2 {\cal M}r-Q^2-P^2)dtd \phi +\frac{\Sigma}{\Delta}dr^2+\Sigma d\theta^2+ \nonumber\\ && \Big(\frac{(2{\cal M}r-Q^2-P^2)a^2\sin^2{\theta}}{\Sigma}-(r^2+a^2)\Big) \sin^2\theta d\phi^2,
\end{eqnarray}
where
\begin{eqnarray} \label{k2}
\Delta&=&r^2-2{\cal M}r+a^2+Q^2+P^2,\nonumber\\
\Sigma&=&r^2+a^2\cos^2{\theta}.
\end{eqnarray}The electromagnetic potential is
\begin{equation}A= \frac{Qr}{\Sigma}(dt-a\sin^2{\theta }d\phi)+\frac{P\cos{\theta}}{\Sigma}\Big[a dt-(r^2+a^2)d\phi \Big]+ \epsilon P d\phi,\end{equation}
here ${\cal M}$ is  mass of the black hole, $Q$ is the electric charge, $P$ is the magnetic charge, $J$ is the angular momentum, and $\epsilon=0, ~\pm 1$. Horizons
of the black hole are obtained by solving $\Delta=0$, given as
\begin{equation} \label{k3}
r_{\pm}={\cal M}\pm \sqrt{{\cal M}^2 -a^2-Q^2-P^2},
\end{equation}
where $r_+$ is the event horizon
${\cal H}^+$ and $r_-$ is the Cauchy
horizon ${\cal H}^{-}$. Both the  Cauchy horizon and the  event horizon are null surfaces with infinite blue and red-shifts respectively \cite{ch83}. Product of two horizons
\begin{equation}\label{k4}
r_+ r_- = a^2+Q^2+P^2,
\end{equation}
does not directly depend on the mass of the black hole but on charge (electric and magnetic) and spin parameter of the black hole.
The area and the Bakenstein-Hawking entropy at horizons ${\cal H}^{\pm}$ of the black hole are \cite{ha4471}
\begin{eqnarray}\label{k5}
\mathcal{A_{\pm}}&=&4\mathcal{S}_{\pm}= \int^{2\pi}_0\int^\pi_0 \sqrt{g_{\theta\theta} g_{\phi \phi}}d\theta d\phi = 4\pi (r_{\pm}^2+a^2)\nonumber\\&&
= 4\pi(2{\cal M}r_{\pm} -Q^2 -P^2),
\end{eqnarray}
The surface gravity is the acceleration due to gravity at the horizon of a black hole. It is defined as the force required to an observer at infinity, for holding a particle (of unit mass) in place at the event horizon, given as \cite{po02}
\begin{eqnarray}\label{k8}\kappa_{\pm}&=&2\pi T_{\pm}=\frac{r_{+}-r_{-}} {2(r_{\pm}^2+a^2)},\nonumber\\
&=&\frac{\sqrt{{\cal M}^2-a^2-P^2-Q^2}}{(r_{\pm}^2+a^2)}.\end{eqnarray}
The Komar energy of the black hole is given by \cite{ko3459}
\begin{eqnarray}
E_{\pm}&=& 2 \mathcal{S}_{\pm} T_{\pm}=\frac{{ r_{+}-r_{-}}}{2},\nonumber\\
&=&\sqrt{{\cal M}^2-a^2-Q^2-P^2}.\label{k10}
\end{eqnarray}
Electromagnetic potentials due to electric and magnetic charges are
\begin{equation}\Phi_{\pm}^Q=\frac{Qr_{\pm}}{r_{\pm}^2+a^2}= \frac{Q r_{\pm}}{2{\cal M}r_{\pm}-Q^2-P^2},\end{equation}
and \begin{equation}\Phi_{\pm}^P=\frac{Pr_{\pm}}{r_{\pm}^2+a^2}= \frac{P r_{\pm}}{2{\cal M}r_{\pm}-Q^2-P^2},\end{equation}
respectively.
Angular velocity at ${\cal H}^{\pm}$ is
\begin{equation}\Omega_{\pm}=\frac{a}{r_{\pm}^2+a^2}=\frac{a}{2{\cal M}r_{\pm}-Q^2-P^2}.\end{equation}
Products of surface gravities, surface temperatures and Komar energies at ${\cal H}^{\pm}$ are
\begin{equation}
\kappa_+\kappa_-=4\pi^2 T_+ T_-= \frac{(r_{\pm}-{\cal M})^2}{[(Q^2+P^2)^2+4a^2{\cal M}^2]},\label{k11}
\end{equation}
and
\begin{eqnarray}\label{k13}
E_+E_-&=&\frac{r_{+}^2-r_{-}^2}{2}, \nonumber\\
&=& {\cal M}^2-a^2-Q^2-P^2,
\end{eqnarray}
respectively.
Products of electromagnetic potentials ${\Phi}$ at ${\cal{H}}^{\pm}$, (due to electric and magnetic charges) are
 \begin{eqnarray}\Phi_{+}^Q\Phi_{-}^Q&=&\frac{Q^2(a^2+Q^2+P^2)}{4{\cal M}^2 a^2+(Q^2+P^2)^2},\nonumber\\ \Phi_{+}^P\Phi_{-}^P&=&\frac{Q^2(a^2+Q^2+P^2)}{4{\cal M}a^2+(Q^2+P^2)^2},\end{eqnarray}
 and product of angular velocities $\Omega$ at ${\cal H}^{\pm}$ is
 \begin{eqnarray}\Omega_+\Omega_-&=&\frac{a^2}{4{\cal M}^2 a^2+(P^2+Q^2)^2},\nonumber\\
 &=& \frac{J^2}{{\cal M}^2(4J^2 +(P^2+Q^2)^2)}.
 \end{eqnarray}It is clear that all these products (except product of horizons $r_+r_-$) are depending on mass of the black hole directly, so these quantities are not universal. We also calculate products of areas
and entropies at ${\cal H}^{\pm}$,
\begin{equation}\mathcal{A}_+\mathcal{A}_-=16 \mathcal{S}_+\mathcal{S}_-= 16 \pi^2[4J^2+(Q^2+P^2)^2].\label{k area}
\end{equation}
 Note that both products are independent of the mass of the black hole, so these are universal quantities.

\section{First Law of Thermodynamics}
Physical parameters defined in previous section for KNK black hole obey symmetries
\begin{eqnarray}\label{cft1} \kappa_+&=&-\kappa_-\mid _{r_-\leftrightarrow {r_+}}, ~~~ {\cal S}_+={\cal S}_- \mid_{r_-\leftrightarrow r_+},~~~ {\Omega}_+={\Omega}_- \mid_{r_-\leftrightarrow {r_+}}, \nonumber\\ &&
\Phi_+=\Phi_- \mid_{r_-\leftrightarrow {r_+}}, ~~~E_+=E_- \mid_{r_-\leftrightarrow {r_+}}. \end{eqnarray}
Solving together $\Delta(r_+)=0=\Delta(r_-)$, black hole parameters ${\cal M}$ and $a$ can be written in terms of horizons $r_{\pm}$, $Q$ and $P$ as
\begin{equation}{\cal M}=\frac{r_++r_-}{2}, ~~~ a^2=r_+r_--Q^2-P^2, \end{equation}
All the parameters studied in previous section can be rewritten as
\begin{eqnarray}\label{k2.5}J&=& \frac{(r_++r_-)\sqrt{r_+r_- -Q^2-P^2}}{2},\nonumber\\&&
\kappa_{\pm}=\frac{r_+-r_-}{2 [r_{\pm}(r_++r_-)-Q^2-P^2]},\nonumber\\&&
{\cal A}_{\pm}= 4\pi [r_{\pm}(r_++r_-)-Q^2-P^2],\nonumber\\&&
\Phi_{\pm}^Q= \frac{Q r_{\pm}}{r_+(r_++r_-)-Q^2-P^2},\nonumber\\&&
\Phi_{\pm}^P= \frac{P r_{\pm}}{r_+(r_++r_-)-Q^2-P^2},\nonumber\\&&
\Omega_{\pm}=\frac{\sqrt{r_+r_--Q^2-P^2}}{r_{\pm}(r_++r_-)-Q^2-P^2}.
\end{eqnarray}
Using the above relations the first law of thermodynamics
\begin{equation}\label{3cft} d{\cal M}= \frac{\kappa_+}{8\pi}d{\cal A}_+ +\Omega_+ d {J}+ \Phi_{+}^Qd Q+\Phi_{+}^P d P,\end{equation} can be verified easily.
The first law at the inner horizon can be written by exchanging $r_+\leftrightarrow r_-$ in all  quantities given in Eq. (\ref{k2.5}), because $r_{\pm}$ are on equal footing. This symmetry of $r_{\pm}$ gives \cite{ca0812, li5410}
\begin{equation}\label{4cft} d{\cal M}= -\frac{\kappa_-}{8\pi}d{\cal A}_-+\Omega_- d {J}+ \Phi_{-}^Q dQ+\Phi_{-}^P d P.\end{equation}
Also it is very obvious that for the  KNK black hole
\begin{equation}T_{+} {\cal S}_{+}= T_{-}{\cal S}_{-}= \frac{r_{+}-r_{-}}{4},\end{equation} products are independent of ${\cal M}$, in agreement with results of work done in  \cite{li5410}, where authors worked for several black holes to show that the mass independence of entropy products together with first laws at both the horizons imply that $T_{+} {\cal S}_{+}=T_{-} {\cal S}_{-}$.
\subsection{The Smarr Formula for the Cauchy Horizon (${\cal H}^-)$}
The expression for the area of the black hole can be rewritten using the idea proposed by the Smarr \cite{sm7173} as:
\begin{equation}\label{k14}
\mathcal{A}=4 \pi \Big[ 2 {\cal M}^2 -Q^2-P^2 + 2{\cal M}\sqrt{{\cal M}^2-a^2-Q^2-P^2}\Big].
\end{equation}
The area of both horizons must be constant given by:
\begin{equation}\label{k15}
\mathcal{A}_{\pm}=4 \pi \Big[ 2 {\cal M}^2 -Q^2-P^2 \pm 2{\cal M}\sqrt{{\cal M}^2-a^2-Q^2-P^2}\Big].
\end{equation}
Using Eq. (\ref{k15}) mass of the black hole or the ADM mass is
expressed in terms of areas of horizons
\begin{eqnarray}\label{M16}
{\cal M}^2&=& \frac{\mathcal{A}_{\pm}}{16 \pi}+\frac{(Q^2+P^2)^2 \pi }{\mathcal{A}_{\pm}}
+\frac{Q^2+P^2}{2}+ \frac{4\pi J^2}{\mathcal{A}_{\pm}}.
\end{eqnarray}
Since surface tension of the black hole is proportional to the temperature of the  black hole, so  the first law of thermodynamics can be written as
\begin{equation}\label{k17}
d{\cal M}=\mathcal{T_{\pm}} d\mathcal{A}_{\pm} +
\Phi_{\pm}^Q dQ+ \Phi_{\pm}^P dP+\Omega_{\pm} dJ,
\end{equation}
where
\begin{eqnarray}\label{M18} \mathcal{T}_{\pm}&=& \text{Effective surface tension at horizons},\nonumber\\
&=&\frac{1}{32 \pi {\cal M}}\Big[1- \frac{16\pi^2((P^2+Q^2)^2 + 4J^2)}{{\cal A}^2_{\pm}}\Big].\nonumber\\
\Phi_{\pm }^Q&=& \text{Electromagnetic potential at horizons due to the  electric charge $Q$},\nonumber\\
&=&\frac{Q r_{\pm}}{2{\cal M}r_{\pm}-Q^2-P^2}.\nonumber\\
\Phi_{\pm }^P&=& \text{Electromagnetic potential at horizons due to the  magnetic charge $P$},\nonumber\\
&=&\frac{P r_{\pm}}{2{\cal M}r_{\pm}-Q^2-P^2}.\nonumber\\
\Omega_{\pm}&=& \text{Angular velocity at horizons},\nonumber\\
&=&\frac{a}{r_{\pm}^2+a^2}.
\end{eqnarray}
The effective surface tension is
\begin{eqnarray}
\mathcal{T}_{\pm}&=& \frac{1}{32 \pi {\cal M}}\Big[1- \frac{16 \pi (\pi(P^2+Q^2)^2  + 4\pi J^2 )}{{\cal A}^2_{\pm}}\Big]
\nonumber\\&&
=\frac{1}{32 \pi {\cal M}}\Big[1-\frac{16\pi {\cal M}^2_{\pm}-{\cal A}_{\pm}-8 \pi (P^2+Q^2)}{{\cal A}}\Big]\nonumber\\&&
=\frac{1}{32 \pi {\cal M}}\Big[1-\frac{(8\pi {\cal M}^2-4 \pi(P^2+Q^2))}{4\pi(r_{\pm}^2+a^2)} \Big]
\nonumber\\&&
=\frac{1}{8 \pi}\Big[\frac{r_{\pm}^2-{\cal M}}{r_{\pm}^2+a^2}\Big]\nonumber\\&&
=\frac{\kappa_{\pm}}{8\pi},
\end{eqnarray}
hence the first law of thermodynamics is derived from the Smarr formula.

\subsection{Christodoulou-Ruffini Mass Formula for the KNK Black Hole}
The mass of a black hole could be increased or decreased
but there is no way to decrease the irreducible mass
${\cal M}_{\text{irr}}$ of a black hole. Also the surface area  ${\cal A}_{\pm}$ of black hole horizon $({\cal H}^{\pm})$ never decreases \cite{ba6173} i.e.
 \begin{eqnarray}
 d{\cal A}_{\pm} \geq 0 \label{arth},
\end{eqnarray} so there is a one to one
connection between the surface area and the irreducible mass of a black
hole.
Christodoulou showed that the irreducible mass of a black hole is proportional to
the square root of its surface area \cite{ch5271}. The irreducible
mass of the KNK black hole expressed in terms of its area is \begin{equation}{\cal M}_{\text{irr}\pm}=\sqrt{\frac{\mathcal{A}_{\pm}}{16 \pi}}= \sqrt{\frac{r^2_{\pm}+a^2}{4}}.
\end{equation}
The irreducible mass defined on the inner and outer horizons is
${\cal M}_{\text{irr}-}$ and ${\cal M}_{\text{irr}+}$
respectively.
The product of the irreducible mass at  horizons
$\mathcal{H^{\pm}}$ is
\begin{eqnarray}\label{k22}
{\cal M}_{\text{irr}+} {\cal M}_{\text{irr}-}&=& \sqrt{\frac{\mathcal{A}_+ \mathcal{A}_-}{(16 \pi)^2}},\nonumber\\
&=& \frac{\sqrt{4 a^2 {\cal M}^2+(P^2+Q^2)^2}}{4},\nonumber\\&&
=\frac{\sqrt{4 J^2 +(P^2+Q^2)^2}}{4}.
\end{eqnarray}
This product is universal because it does not depend directly on the mass of the black hole.
The expression for the rest mass of the rotating
charged black hole given by Christodoulou and Ruffini in
terms of its irreducible mass, the  angular momentum and the charge
is \cite{ch5271}
\begin{equation}
{\cal M}^2= ({\cal M}_{\text{irr}\pm}+ \frac{Q^2}{4{\cal M}_{\text{irr}\pm}})^2
+ \frac{J^2}{4 {\cal M}^2_{\text{irr}\pm}}.\label{M19}
\end{equation}
For the KNK black hole, expression of mass in terms of the irreducible mass is
\begin{equation}{\cal M}^2=\frac{\Big(4{\cal M}_{\text{irr}\pm}^2+(P^2+Q^2)\Big)^2}{(4{\cal M}_{\text{irr}\pm})^2}+\frac{J^2}{4{\cal M}^2_{\text{irr}{\pm}}} ,
\end{equation}
or \begin{equation}{\cal M}^2=\frac{\Big(\rho^2+(P^2+Q^2)\Big)^2}{(2\rho)^2}+\frac{J^2}{\rho^2},\end{equation}
where $\rho_{\pm}= 2 {\cal M}_{\text{irr}_{\pm}}$.

\section{Heat Capacity $C_{\pm}$ on ${\cal H}^{\pm}$}
The nature (positivity or negativity) of the heat capacity reflects the change
in stability properties of the thermal system (black hole). A black hole with negative heat capacity is in unstable equilibrium state i.e. by emitting Hawking radiations it may decay to a hot flat space or by absorbing a radiation it may grow without limit \cite{gr3082}.
The heat capacity of a black hole is given by
\begin{equation}
C_{\pm}= \frac{\partial {\cal M}}{\partial T_{\pm}},
\end{equation}
where mass ${\cal M}$ in terms of $r_{\pm}$ is
\begin{equation}
{\cal M}=\frac{r_{\pm}^2 +a^2 +Q^2+P^2}{2r_{\pm}}.
\end{equation}
Partial derivatives of mass ${\cal M}$ and temperature $T_{\pm}$ with respect to $r_+$ are
\begin{equation}
\frac{\partial {\cal M}}{\partial r_{\pm}}=\frac{r_{\pm}^2-a^2-Q^2-P^2}{4\pi (r_{\pm}^2+a^2)r_{\pm}},
\end{equation}
and
\begin{equation}
\frac{\partial T}{\partial r_{\pm}}=\frac{(a^4-r_{\pm}^4)+(P^2+ Q^2)(3 r_{\pm}^2+a^2)+4a^2 r^2_{\pm}}{4 \pi (r_{\pm}^3+a^2r_{\pm})^2}.
\end{equation}
 \begin{figure}
\centering
\includegraphics [width=9cm]{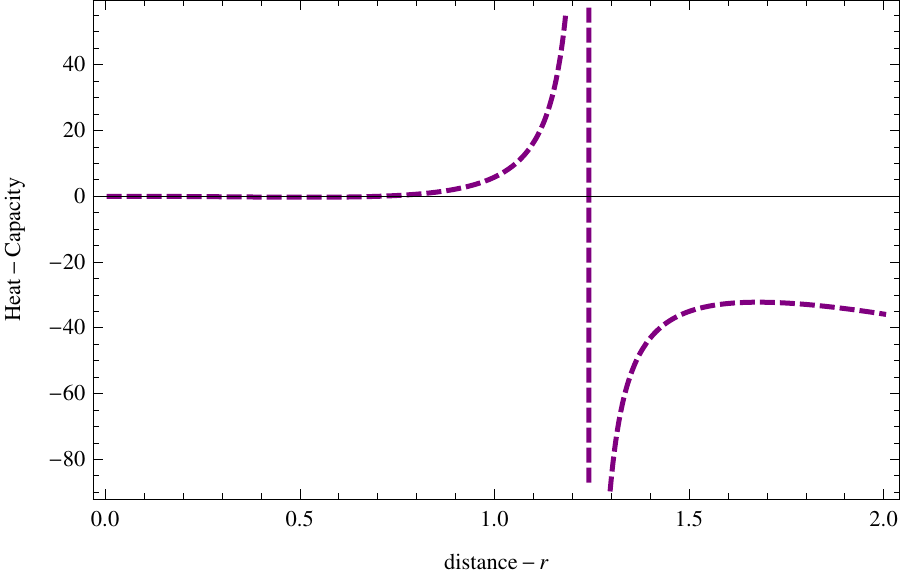}
\caption{Note that Heat capacity is negative for $0<r<0.71414$ and  $1.2423<r<\infty$ while positive for $0.71414<r<1.2423$, so it undergoes phase transition from instable to stable region. Heat capacity diverges at $a^4-r_{\pm}^4+4a^2 r_{\pm}^2+(P^2+Q^2)(3 r_{\pm}^2+a^2)=0$. Here we chose $Q=0.5$, $P=0.5$, $a=0.1$, and divergence occurs at $r= 1.2423$. } \label{1k}\end{figure}
The expression for the heat capacity $C_{\pm}=\frac{\partial {\cal M}}{\partial r_{\pm}}
\frac{\partial r_{\pm}}{\partial T_{\pm}}$ for KNK black hole becomes:
\begin{equation} \label{k21}
C_{\pm}=\Big(\frac{r_{\pm}^2-(a^2+P^2+Q^2)}{2r_{\pm}^2}\Big)\Big(\frac{4 \pi(r_{\pm}^3+a^2 r_{\pm})^2}{a^4-r_{\pm}^4+4a^2r_{\pm}^2+(P^2+Q^2)(3r_{\pm}^2+a^2)}\Big).
\end{equation}
Note that the heat capacity will be positive if\\
 $r_{\pm}^2 -(a^2+P^2+ Q^2)$ and $a^4-r_{\pm}^4+4a^2 r_{\pm}^2+(P^2+Q^2)(3 r_{\pm}^2+a^2)$ are positive or negative together, and heat capacity is negative if these factors are of opposite signs.
The region where the heat capacity is negative, corresponds
to an instable region around black hole, whereas the region in
which the heat capacity is positive, represents a stability
region.
Behavior of the heat capacity given in Eq. (\ref{k21}) is shown in Fig. (\ref{1k}).
\section{The Reissner-Nordstr\"{o}m black hole with the Global Monopole}
A general non-extremal Reissner-Nordstr$\ddot{o}$m black hole with the global
$O(3)$ monopole (RN-GM) has metric \cite{ba4389}\begin{eqnarray}\label{mono1}
ds^2&=& \Big(1-\eta^2-\frac{2{\cal M}}{r}-\frac{Q^2}{r^2}\Big)dt^2-\Big(1-\eta^2-\frac{2{\cal M}}{r}+\frac{Q^2}{r^2}\Big)^{-1}dr^2 - r^2 ( d\theta^2+\sin^2\theta d\phi^2),
\end{eqnarray}
The electromagnetic potential is
\begin{equation}A= \frac{Q}{r}dt,\end{equation}
here ${\cal M}$ is mass of the black hole and $Q$ is the electric charge, $\eta$ is the symmetry
breaking scale for the formation of global monopole, soon after
the Big Bang \cite{pr6879}. The black hole has two  horizons ${\cal H}^{\pm}$, $r_-$ named as the inner (Cauchy) horizon and $r_+$ named as the outer (event) horizon given as
\begin{equation} \label{mono3}
r_{\pm}=\frac{{\cal M}\pm \sqrt{{\cal M}^2 -(1-\eta^2)Q^2}}{(1-\eta^2)}. \end{equation}
Both  horizons give mass independent product
\begin{equation}\label{mono4}
r_+ r_- = \frac{Q^2}{(1-\eta^2)},
\end{equation}
The area and the  entropy of the black hole at ${\cal H}^{\pm}$ are \cite{yu4094}:
\begin{eqnarray}\label{mono5}
\mathcal{A_{\pm}}&=& \int^{2\pi}_0\int^\pi_0 \sqrt{g_{\theta\theta} g_{\phi \phi}}d\theta d\phi = 4\pi (r_{\pm}^2),\nonumber\\&&
= \frac{4\pi(2{\cal M}r_{\pm} -Q^2)}{1-\eta^2},
\end{eqnarray}
and
\begin{eqnarray}\label{mono7}
\mathcal{S}_{\pm}&=&\frac{\mathcal{A}_{\pm}}{4}= \pi (r_{\pm}^2),\nonumber\\ &&
=\frac{\pi(2{\cal M}r_{\pm} -Q^2)}{1-\eta^2}.
\end{eqnarray}
Since the metric given in Eq. (\ref{mono1}) is not asymptotically flat, though asymptotically bounded, the surface gravity of the black hole with this metric can be calculated by using the formula \cite{yu4094}
\begin{eqnarray}\label{mono8}\kappa_{\pm}&=&\frac{\sqrt{1-\eta^2}}{2}(\partial_t g_{tt}\mid_{r=r_{\pm}}),\nonumber\\
&=&\frac{r_{+}-r_{-}} {2r_{\pm}^2 \sqrt{1-\eta^2}}.\end{eqnarray}
The Hawking temperature and the Komar energy \cite{ko3459} of horizons are
\begin{eqnarray}\label{mono9}
T_{\pm}&=&\frac{r_{+}-r_{-}} {4 \pi r_{\pm}^2 \sqrt{1-\eta^2}}, \nonumber\\&&
= \frac{2\sqrt{{\cal M}^2-(1-\eta^2)Q^2}}{4 \pi r_{\pm}^2 \sqrt{1-\eta^2}},
\end{eqnarray}
and
\begin{eqnarray}
E_{\pm}&=&\frac{\sqrt{{\cal M}^2-(1-\eta^2)Q^2}}{(1-\eta^2)^{3/2}},\label{mono10}
\end{eqnarray}respectively.
The Electromagnetic potential caused by the electric  charge is \cite{wu2307}
\begin{equation}\Phi^Q_{\pm}=\frac{Q}{(\sqrt{1-\eta^2})r_{\pm}}=\frac{Q \sqrt{1-\eta^2}}{\sqrt{{\cal M}^2-(1-\eta^2)Q^2}}.\end{equation}
Surface gravities, surface temperatures, Komar energies and the electromagnetic potential ${\Phi^Q}$ at ${\cal{H}}^{\pm}$  give  mass dependent products, so are not universal quantities, these products are
\begin{equation}
\kappa_+\kappa_-= \frac{{\cal M}^2-(1-\eta^2)Q^2}{Q^4(1-\eta^2)},\label{mono11}
\end{equation}
\begin{equation} \label{mono12}
T_+ T_- =\frac{{\cal M}^2-(1-\eta^2)Q^2}{4\pi^2 Q^4(1-\eta^2)},
\end{equation}
\begin{eqnarray}\label{mono13}
E_+E_-&=& \frac{{\cal M}^2-Q^2(1-\eta^2)}{(1-\eta^2)^3},
\end{eqnarray}
and
 \begin{eqnarray}\Phi^Q_{+}\Phi^Q_{-}&=&1,\end{eqnarray} respectively.
From areas
and entropies at ${\cal H}^{\pm}$ we get products depending on mass of the black hole, given as
\begin{equation}\mathcal{A}_+\mathcal{A}_-= \frac{16 \pi^2Q^4}{(1-\eta^2)^2},\label{mono area}
\end{equation}
and
\begin{equation}\label{mono entropy}
\mathcal{S}_+\mathcal{S}_-=\frac{ \pi^2 Q^4}{(1-\eta^2)^2},
\end{equation}
so the area and entropies are universal quantities.
\subsection{First Law of Thermodynamics}Since the entropy product ${\cal S}_+{\cal S}_-$ is mass independent for the RN-GM black hole, which is like $T_+ {\cal S_+} ={T}_-{\cal S_-} $ \cite{li5410} where the product is mass independent given as
\begin{equation}T_{+} {\cal S}_{+}= T_{-}{\cal S}_{-}= \frac{{ r_{+}-r_{-}}}{4 \sqrt{1-\eta^2}}.\end{equation}
Also thermodynamics parameters defined for the RN-GM black hole obey  symmetries
\begin{eqnarray}\label{monocft1} \kappa_+&=&-\kappa_-\mid _{r_-\leftrightarrow {r_+}}, ~~~ {\cal S}_+={\cal S}_- \mid_{r_-\leftrightarrow r_+}, \nonumber\\ &&
\Phi^Q_+=\Phi^Q_- \mid_{r_-\leftrightarrow {r_+}}, ~~~E_+=E_- \mid_{r_-\leftrightarrow {r_+}}. \end{eqnarray}
We can write the first law of thermodynamics
\begin{equation}\label{3monocft} d{\cal M}= \frac{\kappa_+}{8\pi}d{\cal A}_++ \Phi^Q_{+}d Q,\end{equation}at the inner horizon as well by exchanging $r_+\leftrightarrow r_-$ in all quantities of black hole involved  and collectively  one can write \cite{ca0812, li5410}
\begin{eqnarray}\label{4monocft}  d{\cal M}&=&\frac{\kappa_+}{8\pi}d{\cal A}_++ \Phi^Q_{+}d Q, \nonumber\\
&=&-\frac{\kappa_-}{8\pi}d{\cal A}_-+\Phi^Q_{-} dQ
\end{eqnarray}
Central charges and temperatures of holographic  pictures of the black hole can be gained by applying the thermodynamics approach \cite{cv0909}.
\subsection{The Smarr Formula for the Cauchy Horizon (${\cal H}^-)$}
The area of  the RN-GM black hole horizons is
\begin{equation}\label{mono15}
\mathcal{A}_{\pm}=\frac{4\pi}{(1-\eta^2)^2} \Big[ 2 {\cal M}^2 -(1-\eta^2) Q^2 \pm 2{\cal M}\sqrt{{\cal M}^2-Q^2(1-\eta^2)}\Big].
\end{equation}
Using Eq. (\ref{mono15}), the mass of the black hole or the ADM mass can be
expressed in terms of areas of horizons as
\begin{eqnarray}\label{monoM16}
{\cal M}^2&=& \frac{\mathcal{A}_{\pm}(1-\eta^2)^2}{16 \pi}+\frac{(1-\eta^2) Q^2}{2}+\frac{Q^4\pi}{\mathcal{A}_{\pm}(1-\eta^2)}.
\end{eqnarray}
 We can write the change in the mass of the black hole, in terms of change in the area and change in the electric charge (known as the first law of thermodynamics), i.e.
\begin{equation}\label{monok17}
d{\cal M}=\mathcal{T_{\pm}} d\mathcal{A}_{\pm} +
\Phi^Q_{\pm} dQ,
\end{equation}
where
\begin{eqnarray}\label{M18} \mathcal{T}_{\pm}&=& \text{Effective surface tension at horizons},\nonumber\\
&=&\frac{(1-\eta^2)^2}{32 \pi {\cal M}}\Big[1- \frac{16\pi^2 Q^4}{(1-\eta^2)^3{\cal A}^2_{\pm}}\Big]\nonumber\\
\Phi^Q_{\pm }&=&\frac{Q(1-\eta^2)^2+4\pi Q^3}{2{\cal M}{\cal A}_{\pm} (1-\eta^2)}.
\end{eqnarray}
We can rewrite effective surface tension as
\begin{eqnarray}
\mathcal{T}_{\pm}&=& \frac{(1-\eta^2)^2}{32 \pi {\cal M}}\Big[1- \frac{16 \pi {\cal M}^2-{\cal A}_{\pm}(1-\eta^2)^2-8 \pi (1-\eta^2) Q^2}{(1-\eta^2){\cal A}_{\pm}}\Big],
\nonumber\\&&
=\frac{1}{16 \pi {\cal M}}\Big[\frac{(1-\eta^2)^2 r_{\pm}^2 -2{\cal M}^2+(1-\eta^2)Q^2}{r_{\pm}^2} \Big],
\nonumber\\&&
=\frac{1}{8 \pi r_{\pm}^2}\Big[r_{\pm}(1-\eta^2)-{\cal M}\Big],\nonumber\\&&
=\frac{\kappa_{\pm}}{8\pi},
\end{eqnarray}
hence the first law of thermodynamics is derived from the Smarr formula approach.

\section{Conclusion}Some important thermodynamical parameters for the Kerr-Newman-Kasuya black hole
with reference to their event and the  Cauchy horizon  are studied. We find products of these parameters calculated at both horizons of the black hole, and observe that the product of
surface gravities, surface temperature product, product of
Komar energies, products of electromagnetic potentials and angular velocities at  horizons are not universal quantities, being mass dependant, while products of areas and
entropies at both horizons are independent of mass of the black hole.
We derive the expression for heat capacity at horizons.
Using the heat capacity expression, stability regions of the black hole
 can be estimated and it is observed graphically for KNK black hole.  The first law of thermodynamics is also obtained
  from the Smarr formula approach.
Relations that are independent of the
black hole mass are of particular interest because these may turn out to be ``universal'' and  hold
for more general solutions, with nontrivial surroundings. The relation ${T}_+{\cal S_+} = {T}_-{\cal S_-} $ holds for the KNK  black hole, so it has a CFT dual in the Einstein gravity and KNK/CFT correspondence is achievable.
Thermodynamics aspects of the RN black hole with global monopole (RN-GM) are also extended to the inner (Cauchy) horizon of the black hole.
  Products of thermodynamics parameters (calculated at both horizons of the black hole)  show that the product of
surface gravities (and so is surface temperature product), product of
Komar energies and products of electromagnetic potentials are mass dependant, while products of areas (and so products of
entropies) are not depending on mass of the black hole so these are universal quantities. Also mass of the black hole is calculated in terms of its irreducible mass.\\
 Since for  both black holes discussed in this work, ${T}_+{\cal S_+} = {T}_-{\cal S_-} $ holds so a RN-GM/CFT correspondence can be established for these black holes. The
temperature of  CFT duals and  central charges
can be obtained by applying the quantization rule on the perturbation of the black holes (the first law of thermodynamics used in terms of quantized charges), this approach of work will be adopted in some later work.


\end{document}